\documentclass[11pt,fleqn]{article}
\usepackage{amsmath,amssymb,graphicx}

\begin{document}
\sloppy
\newtheorem{axiom}{Axiom}[section]
\newtheorem{conjecture}[axiom]{Conjecture}
\newtheorem{corollary}[axiom]{Corollary}
\newtheorem{definition}[axiom]{Definition}
\newtheorem{example}[axiom]{Example}
\newtheorem{fact}[axiom]{Fact}
\newtheorem{lemma}[axiom]{Lemma}
\newtheorem{observation}[axiom]{Observation}
\newtheorem{open}[axiom]{Problem}
\newtheorem{proposition}[axiom]{Proposition}
\newtheorem{theorem}[axiom]{Theorem}

\renewcommand{\topfraction}{1.0}
\renewcommand{\bottomfraction}{1.0}

\newcommand{\proof}{\emph{Proof.}\ \ }
\newcommand{\qed}{~~$\Box$}
\newcommand{\rz}{{\mathbb{R}}}
\newcommand{\nz}{{\mathbb{N}}}
\newcommand{\zz}{{\mathbb{Z}}}
\newcommand{\eps}{\varepsilon}
\newcommand{\cei}[1]{\lceil #1\rceil}
\newcommand{\flo}[1]{\left\lfloor #1\right\rfloor}

\newcommand{\aaa}{\alpha}
\newcommand{\bbb}{\beta}
\newcommand{\ccc}{\gamma}
\newcommand{\ppp}{\sigma}

\newcommand{\ddd}[1]{d(#1)}
\newcommand{\kalm}[1]{\mbox{$#1$-Kalmanson}}
\newcommand{\cromat}{\mbox{\sc CrossMatch}}

\title{{\bf The multi-stripe travelling salesman problem}}
\author{
Eranda \c{C}ela\thanks{{\tt cela@opt.math.tu-graz.ac.at}.
Institut f\"ur Optimierung und Diskrete Mathematik, TU Graz, Steyrergasse 30, A-8010 Graz, Austria}
\\Vladimir G.\ Deineko\thanks{{\tt Vladimir.Deineko@wbs.ac.uk}.
Warwick Business School, The University of Warwick, Coventry CV4 7AL, United Kingdom}
\\Gerhard J.\ Woeginger\thanks{{\tt gwoegi@win.tue.nl}.
Department of Mathematics and Computer Science, TU Eindhoven, P.O.\ Box 513,
5600 MB Eindhoven, Netherlands}
}
\date{}
\maketitle

\begin{abstract}
In the classical Travelling Salesman Problem (TSP), the objective function sums the costs for travelling 
from one city to the next city along the tour.
In the $q$-stripe TSP with $q\ge1$, the objective function sums the costs for travelling from one city to 
each of the next $q$ cities along the tour.
The resulting $q$-stripe TSP generalizes the TSP and forms a special case of the quadratic assignment problem.

We analyze the computational complexity of the $q$-stripe TSP for various classes of specially structured
distance matrices.
We derive NP-hardness results as well as polyomially solvable cases. 
One of our main results generalizes a well-known theorem of Kalmanson from the classical TSP to the
$q$-stripe TSP.

\end{abstract}

\medskip
\section{Introduction}
We consider a generalization of the classical travelling salesman problem (TSP).
Recall that an instance of the TSP consists of $n$ cities together with an $n\times n$ matrix $D=(d_{ij})$ 
that specifies the distances between these cities.
A feasible solution for the TSP is a permutation $\pi\in S_n$ of the cities (also called \emph{tour} 
or \emph{round trip}).
The goal is to find a feasible solution $\pi=\langle\pi(1),\ldots,\pi(n)\rangle$ that minimizes the 
total travel length 
\begin{equation}
\label{eq:tsp}
\mbox{TSP}(\pi) ~=~ \sum_{i=1}^n d(\pi(i),\pi(i+1)).
\end{equation}
(When we do arithmetics with cities, we usually identify city $x$ with the cities $x+n$ and $x-n$; hence 
the term $\pi(i+1)$ for $i=n$ in the preceding formula coincides with $\pi(1)$, so that the salesman in 
the very end returns to the city from which he started his trip.)
We refer the reader to the book \cite{TSP-book} by Lawler, Lenstra, Rinnooy Kan \& Shmoys for a wealth of 
information on the TSP, and to the papers by Gilmore, Lawler \& Shmoys \cite{GLS1985} and 
Burkard \& al \cite{BDDVW1998} for comprehensive surveys on tractable special cases.

An instance of the \emph{$q$-stripe Travelling Salesman Problem} with $1\le q\le (n-1)/2$ consists of $n$ 
cities together with an $n\times n$ distance matrix $D=(d_{ij})$, exactly as in the standard TSP.
The goal is to find a permutation $\pi=\langle\pi(1),\pi(2),\ldots,\pi(n)\rangle$ through all cities 
that minimizes the cost function 
\begin{equation}
\label{eq:qtsp}
\mbox{\sc $q$-Stripe-TSP}(\pi) ~=~ \sum_{p=1}^q\sum_{i=1}^n d(\pi(i),\pi(i+p)).
\end{equation}
As for $q=1$ the expression in \eqref{eq:qtsp} coincides with the expression in \eqref{eq:tsp},
the $q$-stripe TSP properly generalizes the classical TSP.
Intuitively speaking, the permutation $\pi$ encodes a tour through the cities, and the expression 
in \eqref{eq:qtsp} sums the distances $d(i,j)$ over all cities $i$ and $j$ that are at most $q$
steps away from each other when travelling along the tour.

The city pairs $(i,j)$ that contribute to the objective functions \eqref{eq:tsp} and \eqref{eq:qtsp} 
determine the edges of an underlying graph.
For $q=1$ this graph is simply the Hamiltonian cycle $C_n$ on $n$ vertices, and for $q\ge2$ it is 
the $q$-th power of $C_n$ (the graph that results from the cycle $C_n$ by connecting all vertex pairs
that are separated by at most $q$ edges along the cycle).
These observations already indicate a close connection between the $q$-stripe TSP and certain 
graph-theoretic questions, which will be discussed in Section~\ref{sec:literature}.
Furthermore, we will study the computational complexity of a graph-theoretic version that constitutes
a highly structured special case of the $q$-stripe TSP.
In Section~\ref{sec:hard} we will show that the graph-theoretic version is NP-hard in multi-partite graphs 
with $p\ge q+1$ parts, in split graphs, and in graphs that do not contain $K_{1,4}$ as induced sub-graph.
In Section~\ref{sec:easy} we will show that the graph-theoretic version is polynomially solvable in
planar graphs (if $q\ge2$) and in partial $k$-trees (if the parameter $k$ is a fixed constant).

The $q$-stripe TSP may also be interpreted as a special case of the quadratic assignment problem (QAP).
This will be discussed in Section~\ref{sec:literature}, where we also survey the underlying literature
and some consequences for tractable special cases of the $q$-stripe TSP.
Our main result generalizes a tractable special case of the TSP and QAP (formulated in 
Theorems~\ref{th:Kalmanson} and~\ref{th:Dei+Woe}) from the class of so-called Kalmanson matrices to 
a broader class of matrices that we call $q$-Kalmanson matrices; see Section~\ref{sec:Kalmanson}.
As a by-product, we derive in Section~\ref{sec:master} a complete characterization of the distance 
matrices that allow a so-called master $q$-stripe TSP tour; a master tour simultaneously induces 
optimal solutions to all possible sub-instances of a given problem instance.
Section~\ref{sec:discussion} completes the paper with a discussion and several open questions.

\section{Technical preliminaries and literature review}
\label{sec:literature}
In this section, we survey results from graph theory and from combinatorial optimization that yield 
tractable special cases of the $q$-stripe TSP.
We also introduce a number of definitions that will be crucial in the rest of the paper.

A graph-theoretic version of the $q$-stripe TSP is centered around the $q$-th power of the undirected 
cycle $C_n$ on $n$ vertices: 
For $q\ge1$ and $n\ge2q+1$ the vertex set of graph $C^q_n$ is $\{1,2,\ldots,n\}$, and there is an edge 
between any two distinct vertices $i$ and $j$ with $|i-j|\le q$ or $|i-j|\ge n-q$.
For $q=1$, the resulting graph $C^1_n$ coincides with the standard undirected $n$-vertex cycle $C_n$.
Note that the graph $C^q_n$ encodes the cost structure of the $q$-stripe TSP on $n$ cities.
Furthermore, the problem of finding a spanning sub-graph $C^q_n$ in a given input graph $G$ is a special
case of the $q$-stripe TSP: for any edge $[i,j]$ in $G$ we set $d(i,j)=0$, for any non-edge we set $d(i,j)=1$,
and we ask for a permutation $\pi$ for which the objective value \eqref{eq:qtsp} is $0$.

Paul Seymour \cite{Seymour1973} conjectured that every $n$-vertex graph with minimum degree at least $qn/(q+1)$ 
contains a spanning sub-graph $C^q_n$.
Koml\'os, S\'ark\"ozy \& Szemer\'edi \cite{KoSaSz1998} proved this conjecture for sufficiently large $n$
by using Szemer\'edi's regularity lemma.
Donnelly \& Isaak \cite{DoIs1999} present a variety of combinatorial and algorithmic results on spanning
sub-graphs $C^q_n$ in threshold graphs (graphs that do not contain $C_4$, $P_4$, $2K_2$ as induced sub-graph)
and in arborescent comparability graphs (graphs that do not contain $C_4$ or $P_4$ as induced sub-graph);
in particular, they design polynomial time algorithms for detecting such spanning sub-graphs $C^q_n$ for
these classes.
The complexity of detecting a spanning $C^q_n$ in interval graphs is an open problem; see Isaak \cite{Isaak1998}.

The $q$-stripe TSP may also be formulated in a natural way as a quadratic assignment problem (QAP) in 
Koopmans-Beckmann form \cite{KoBe1957}. 
The QAP takes as input two $n\times n$ matrices $D=(d(i,j))$ and $C=(c(i,j))$, and assigns to 
permutation $\pi=\langle\pi(1),\pi(2),\ldots,\pi(n)\rangle$ a corresponding objective value
\begin{equation}
\label{eq:qap}
\mbox{QAP}(\pi) ~:=~ \sum_{i=1}^n\sum_{j=1}^n~ \ddd{\pi(i),\pi(j)} \cdot c(i,j).
\end{equation}
The goal is to find a permutation $\pi$ that minimizes the objective value.
By making matrix $D$ the distance matrix of $n$ cities and by making matrix $C=(c(i,j))$ the adjacency matrix 
of the graph $C^q_n$, we arrive at the $q$-stripe TSP as a special case of the QAP.  
We refer the reader to the books by Burkard, Dell'Amico \& Martello \cite{Burkard-book} and 
\c{C}ela \cite{Cela-book} for detailed information on the QAP.
In particular, the QAP literature contains a number of tractable special cases that are built around certain
combinatorial structures in the underlying cost matrices.
We will discuss some of these special cases in the following paragraphs and relate them to the $q$-stripe TSP.

An $n\times n$ matrix $D$ is a \emph{Monge matrix} if its entries fulfill the following conditions \eqref{eq:Monge}.
\begin{equation}
\label{eq:Monge}
\ddd{i,j}+\ddd{r,s} ~\le~ \ddd{i,s}+\ddd{r,j} \mbox{\qquad for $1\le i<r\le n$ and $1\le j<s\le n$.}
\end{equation}
These inequalities (\ref{eq:Monge}) go back to the 18th century, to the work of the French mathematician 
and Naval minister Gaspard Monge \cite{Monge}.
Burkard, Klinz \& Rudolf \cite{BuKlRu1996} survey the important role of Monge structures in combinatorial 
optimization, and summarize the vast literature.
Fred Supnick \cite{Supnick1957} proved in 1957 by means of an exchange argument that for the TSP on symmetric 
Monge matrices, an optimal tour is easy to find and in fact is always given by a fixed permutation $\ppp$. 
\begin{theorem}
\label{th:Supnick}
(Supnick \cite{Supnick1957})
The TSP on symmetric Monge matrices is solvable in polynomial time.
A shortest TSP tour is given by the permutation
\begin{equation}
\label{eq:Supnick}
\ppp=\langle 1,3,5,7,9,11,13,\ldots14,12,10,8,6,4,2\rangle.
\end{equation}
This permutation $\ppp$ first traverses the odd cities in increasing order and then traverses the even 
cities in decreasing order.
\end{theorem}
Burkard, \c{C}ela, Rote \& Woeginger \cite{BCRW1998} generalized and unified several special cases of the QAP.
Their main result implies the following generalization of Supnick's result to the $q$-stripe TSP.
\begin{theorem}
\label{th:Monge}
(Burkard, \c{C}ela, Rote \& Woeginger \cite{BCRW1998})
For any $q\ge1$, the $q$-stripe TSP on symmetric Monge matrices is solvable in polynomial time.
The permutation $\ppp$ in \eqref{eq:Supnick} always yields an optimal solution.
\end{theorem}

An $n\times n$ symmetric matrix $D$ is a \emph{Kalmanson matrix} if its
entries fulfill the following two families of conditions:
\begin{eqnarray}
\ddd{i,j}+\ddd{k,\ell} ~\le~ \ddd{i,k}+\ddd{j,\ell} && \mbox{\quad for all $1\le i<j<k<\ell\le n$} \label{ka.1} \\
\ddd{i,\ell}+\ddd{j,k} ~\le~ \ddd{i,k}+\ddd{j,\ell} && \mbox{\quad for all $1\le i<j<k<\ell\le n$} \label{ka.2}
\end{eqnarray}
Kalmanson matrices were introduced by Kalmanson \cite{Kalmanson1975} in his investigations of special 
cases of the travelling salesman problem.
They form a common generalization of the following two well-known families of distance matrices.
First, the distance matrix of every convex point set in the Euclidean plane forms a Kalmanson matrix, 
if the points are numbered in (say) clockwise direction along the convex hull.  
The inequalities (\ref{ka.1}) and (\ref{ka.2}) then simply state that in a convex quadrangle, the
total length of two opposing sides is at most the total length of the two diagonals.
Secondly, so-called tree metrics correspond to Kalmanson matrices.
Consider a rooted ordered tree with non-negative edge lengths, and number its leaves from left to right.
Then the shortest path distances between leaves $i$ and $j$ determine a Kalmanson matrix.
Indeed, the inequalities (\ref{ka.1}) and (\ref{ka.2}) are easily verified for sub-trees with four leaves.

Kalmanson matrices play a prominent role in combinatorial optimization.
Bandelt \& Dress \cite{BaDr1992}, and independently Christopher, Farach \& Trick \cite{ChFaTr1996} and 
Chepoi \& Fichet \cite{ChFi1998} showed that the Kalmanson conditions are equivalent to so-called 
\emph{circular decomposable metrics}.
Klinz \& Woeginger \cite{KlWo1999} analyzed the Steiner tree problem in Kalmanson matrices, 
and Polyakovskiy, Spieksma \& Woeginger \cite{PoSpWo2013} investigated a special case of the 
three-dimensional matching problem in Kalmanson matrices.
Kalmanson's paper \cite{Kalmanson1975} contains the following result for the TSP.
\begin{theorem}
\label{th:Kalmanson}
(Kalmanson \cite{Kalmanson1975})
The TSP on Kalmanson distance matrices is solvable in polynomial time.
The identity permutation yields an optimal solution.
\end{theorem}
Deineko \& Woeginger \cite{DeWo1998} generalized Kalmanson's result \cite{Kalmanson1975} to certain 
special cases of the QAP; in particular, their results imply the following theorem for the $q$-stripe TSP.
\begin{theorem}
\label{th:Dei+Woe}
(Deineko \& Woeginger \cite{DeWo1998}) 
For any $q\ge1$, the $q$-stripe TSP on Kalmanson matrices is solvable in polynomial time.
The identity permutation always yields an optimal solution.
\end{theorem}
Finally, we mention the quadratic travelling salesman problem as studied by Fischer \& Helmberg \cite{FiHe2013}.
In this variant, a cost $c(i,j,k)$ is associated with any three cities $i,j,k$ that the salesman traverses 
in succession.
Fischer \& Helmberg argue that this variant arises if the succession of two edges represents energetic 
conformations, a change of direction or a possible change of transportation means.
By setting $c(i,j,k)=\frac12d(i,j)+\frac12d(j,k)+d(i,k)$, we see that the quadratic TSP properly 
generalizes the $2$-stripe TSP.

\section{The $q$-stripe TSP on $q$-Kalmanson matrices}
\label{sec:Kalmanson}
In this section, we will generalize Theorems~\ref{th:Kalmanson} and~\ref{th:Dei+Woe} to a much broader class
of matrices.
For four cities $i<j<k<\ell$, the two edges $[i,k]$ and $[j,\ell]$ are said to be \emph{crossing}.
For an even number of cities $i_1<i_2<\cdots<i_{2k}$, their \emph{fully crossing matching} consists of the 
$k$ edges $[i_j,i_{j+k}]$ with $j=1,\ldots,k$. 
In other words, the fully crossing matching pairs every city in $\{i_1,\ldots,i_{2k}\}$ with its 
diametrically opposed city in the natural circular arrangement of the cities, so that every pair
of edges in this matching is crossing.
The total length of all edges in the fully crossing matching is denoted by $\cromat(\{i_1,\ldots,i_{2k}\})$.

\begin{definition}
\label{df:q-Kalm}
Let $D$ be a symmetric $n\times n$ distance matrix.
A subset of cities satisfies the \emph{\kalm{q} condition}, if the fully crossing matching 
forms a perfect matching of maximum weight on these cities.
Matrix $D$ is said to be a \emph{\kalm{q} matrix}, if every subset of $2q+2$ cities satisfies the 
\kalm{q} condition.
\end{definition}
Note that the \kalm{1} condition coincides with conditions (\ref{ka.1}) and (\ref{ka.2}) as introduced 
in the original paper by Kalmanson \cite{Kalmanson1975}; in other words, the \kalm{1} matrices are exactly 
the standard Kalmanson matrices from the literature.

\begin{lemma}
\label{le:q-Kalm}
For every integer $q\ge1$, the \kalm{q} matrices form a proper subclass of the \kalm{(q+1)} matrices.
\end{lemma}
\proof
Consider $2q+4$ cities $1,2,\ldots,2q+4$ that satisfy the \kalm{q} condition for some distance matrix $D$.
Let $\cal M$ be a maximum weight matching for these cities, and let $[1,x]$ denote the edge that covers 
city $1$ in $\cal M$.
By symmetry we may assume $x\le q+2$, and by the \kalm{q} condition we may assume that the induced 
matching for $\{1,\ldots,2q+4\}\setminus\{1,x\}$ is fully crossing.
If $x\ne2$, then $\cal M$ contains the edge $[2,q+3]$.
In this case we replace the matching on $\{1,\ldots,2q+4\}\setminus\{2,q+3\}$ by the corresponding 
fully crossing matching.
The resulting matching is fully crossing on $1,2,\ldots,2q+4$ and has maximum weight.
If $x=2$, then $\cal M$ contains the edge $[3,q+3]$.
In this case we shift the numbering of cities by $-2$, so that the edge $[3,q+3]$ becomes $[1,q+1]$ and 
apply the above argument to the renumbered instance.
In either case, we see that the $2q+4$ cities satisfy the \kalm{(q+1)} condition.
This settles the subset relation stated in the lemma.

To see that the subset relation between the two matrix classes is proper, we introduce the following 
symmetric $n\times n$ matrix $D_{n,q}$ for $q\ge1$ and $n\ge2q+4$: 
\begin{equation}
\label{eq:example}
d(i,j) ~=~ \left\{ \begin{array}{cl}
        1 & \mbox{~ if $q+2\le|i-j|\le n-q-2$} \\[0.3ex]
        0 & \mbox{~ otherwise}
        \end{array}  \right. 
\end{equation}
Now consider $2q+4$ arbitrary cities $i_1<i_2<\cdots<i_{2q+4}$, and let $[i_j,i_{j+q+2}]$ with $1\le j\le q+2$ 
be an edge in their fully crossing matching.
Then $q+2\le |i_j-i_{j+q+2}|\le n-q-2$, as the $q+1$ cities $i_{j+1},\ldots,i_{j+q+1}$ lie in the interval 
between $i_j$ and $i_{j+q+2}$ whereas the $q+1$ cities $i_1,\ldots,i_{j-1}$ and $i_{j+q+3},\ldots,i_{2q+4}$ 
lie outside this interval.
This means that all edges in the fully crossing matching have weight~$1$, and that the fully crossing
matching indeed is a maximum weight matching.
Therefore $D_{n,q}$ is a \kalm{(q+1)} matrix.  
On the other hand, the fully crossing matching for the first $2q+2$ cities $1,2,\ldots,2q+2$ has weight~$0$.
The matching that consists of edge $[1,q+3]$ of weight~$1$ together with some $q$ other edges has strictly
positive weight.
Therefore $D_{n,q}$ is not a \kalm{q} matrix.
\qed

\bigskip
In the remainder of this section, we will analyze the $q$-stripe TSP on $q$-Kalmanson matrices.
We start with the analysis of an auxiliary optimization problem.
For some fixed city $x$, we are now looking for $2q$ pairwise distinct cities $y_1,y_2,\ldots,y_{2q}$ that all 
are distinct from $x$ and that minimize the objective function
\begin{equation}
\label{eq:reni.0}
f_x(y_1,\ldots,y_{2q}) ~=~ \sum_{i=1}^{2q}d(x,y_i) - \cromat(\{y_1,\ldots,y_{2q}\}).
\end{equation}
The following result will be useful in our investigations.
\begin{lemma}
\label{le:reni}
Let $q\ge1$ and $n\ge2q+1$, and let $D$ be a \kalm{q} matrix.
Then for every city $x$ the function $f_x$ in \eqref{eq:reni.0} is minimized by setting $y_i=x-q+i-1$ 
for $i=1,\ldots,q$ and by setting $y_i=x-q+i$ for $i=q+1,\ldots,2q$.
(In other words, there exists a minimizer that uses the $q$ cities directly preceding $x$ and 
the $q$ cities directly succeeding $x$ in the underlying circular arrangement.) 
\end{lemma}
\proof
Without loss of generality we assume $x=q+1$.
Among all the minimizers $Y=\{y_1,\ldots,y_{2q}\}$ of the function $f_x$, we consider one that secondarily 
maximizes the number of common elements of $Y\cup\{x\}$ and $T=\{1,2,\ldots,2q+1\}$.
Suppose for the sake of contradiction that $Y\cup\{x\}\ne T$, and let $z$ be a city in $T\setminus(Y\cup\{x\})$.
As the distance matrix $D$ satisfies the \emph{\kalm{q}} condition for the $2q+2$ cities in $Y\cup\{x,z\}$, 
we have
\begin{equation}
\label{eq:reni.1a}
\cromat(Y)+d(x,z) ~\le~ \cromat(Y\cup\{x,z\}).
\end{equation}
As $|x-t|\le q$ holds for all $t\in T$, the fully crossing matching for the $2q+2$ cities in $Y\cup\{x,z\}$ 
will match city $x$ with some city $y_j\in Y\setminus T$ (and hence will not match $x$ with $z$).
This yields
\begin{equation}
\label{eq:reni.1b}
\cromat(Y\cup\{x,z\}) ~=~ \cromat(\{z\}\cup Y\setminus\{y_j\})+d(x,y_j).
\end{equation}
Finally we derive from \eqref{eq:reni.0} by using \eqref{eq:reni.1a} and \eqref{eq:reni.1b} that
\begin{eqnarray*}
f_x(Y) &=& \sum_{y\in Y}d(x,y) - \cromat(Y) 
\\[0.5ex]&\ge&
\sum_{y\in Y}d(x,y)+d(x,z) - \cromat(Y\cup\{x,z\})
\\[0.5ex]&=&
\sum_{y\in Y\cup\{z\}}d(x,y) - \cromat(\{z\}\cup Y\setminus\{y_j\})-d(x,y_j)
\\[0.5ex]&=& f_x(\{z\}\cup Y\setminus\{y_j\}).
\end{eqnarray*}
As $z\in T$ and $y_j\notin T$, the set $\{z\}\cup Y\setminus\{y_j\}$ has more elements in common with $T$
than set $Y$, while its objective value is at least as good as the objective value of $Y$. 
That's the desired contradiction.
\qed

\bigskip
The following theorem states our main result on \kalm{q} matrices. 
The rest of this section will be dedicated to its proof.
\begin{theorem}
\label{th:vlad}
For every integer $q\ge1$, the $q$-stripe TSP on a \kalm{q} matrix is solved to optimality by the 
identity permutation $\pi=\langle1,\ldots,n\rangle$.
\end{theorem}
\proof
The proof of the theorem proceeds by induction on the number $n\ge2q+1$ of cities.
For $n=2q+1$, the objective function in \eqref{eq:qtsp} simply adds up the lengths of all the edges
between pairs of distinct cities.
Hence in this case every permutation $\pi\in S_n$ yields the same objective value, and the statement holds trivially.

In the inductive step from $n-1$ to $n$, we consider an arbitrary \kalm{q} distance matrix for 
$n$ cities and an optimal permutation $\pi\in S_n$ for the $q$-stripe TSP.
Without loss of generality we assume $\pi(n)=n$, so that $\pi(1),\pi(2),\ldots,\pi(n-1)$ is a permutation 
of the cities $1,2,\ldots,n-1$.
The inductive assumption yields for the induced instance on the first $n-1$ cities that
\begin{equation}
\label{eq:vlad.1}
\sum_{p=1}^q\sum_{i=1}^{n-1} d(i,i+p) ~\le~ \sum_{p=1}^q\sum_{i=1}^{n-1} d(\pi(i),\pi(i+p)).
\end{equation}
(In this equation arithmetics with cities is done modulo the number $n-1$ of cities, 
so that $x$ coincides with $x+n-1$ and $x-n+1$.)
The $q$ immediate successors of city $n=\pi(n)$ in the tour $\pi$ are $\pi(1),\ldots,\pi(q)$,
and its $q$ immediate predecessors are $\pi(n-q),\ldots,\pi(n-1)$.
Lemma~\ref{le:reni} yields for $x:=n$ that 
\begin{eqnarray}
\lefteqn{\sum_{i=n-q}^{n-1}d(n,i)+\sum_{i=1}^qd(n,i) - \cromat(\{1,\ldots,q\}\cup\{n-q,\ldots,n-1\})}
\nonumber\\[0.5ex] &\le& \sum_{i=n-q}^{n-1}d(n,\pi(i))+\sum_{i=1}^qd(n,\pi(i))
\nonumber\\ && \qquad -~ \cromat(\{\pi(1),\ldots,\pi(q)\}\cup\{\pi(n-q),\ldots,\pi(n-1)\})
\label{eq:vlad.2}
\end{eqnarray}
By adding up the inequalities in \eqref{eq:vlad.1} and \eqref{eq:vlad.2} we get
the desired statement 
\begin{equation}
\label{eq:vlad.3}
\sum_{p=1}^q\sum_{i=1}^{n} d(i,i+p) ~\le~ \sum_{p=1}^q\sum_{i=1}^{n} d(\pi(i),\pi(i+p)).
\end{equation}
Hence, the identity permutation indeed yields the smallest possible objective value for the $q$-stripe TSP.
This completes the proof of Theorem~\ref{th:vlad}.
\qed

\section{Master tours for the $q$-stripe TSP}
\label{sec:master}
Assume that the cities in a Euclidean instance of the TSP are the vertices of a convex polygon.
Then an optimal tour is not only easy to find (it follows the perimeter of the polygon), but the 
instance also possesses a so-called \emph{master tour}:
There exists an optimal TSP tour $\pi$ that simultaneously encodes the optimal tours for all subsets 
of the cities, as an optimal tour for a subset may be obtained by simply omitting from the tour $\pi$ 
all the cities that are not in the subset.
The concept of such master tours was introduced by Papadimitriou \cite{Papadimitriou1993,Papadimitriou1994}. 
Deineko, Rudolf \& Woeginger \cite{DeRuWo1998} showed that a TSP instance has a master tour 
if and only if the underlying distance matrix is a Kalmanson matrix.
Van Ee \& Sitters \cite{EeSi2015} investigate master versions of the Steiner tree problem and of the
maximum weighted satisfiability problem.

In this spirit, let us say that a distance matrix $D$ has a master tour $\pi$ for the $q$-stripe TSP, 
if for any subset $S$ of the cities an optimal $q$-stripe tour can be obtained by removing from $\pi$ 
the cities not contained in $S$.
The following theorem fully settles the combinatorics of master tours with respect to the $q$-stripe TSP.
\begin{theorem}
\label{th:master}
For any $q\ge1$ and for any $n\times n$ distance matrix $D$, the identity permutation is a 
master tour for the $q$-stripe TSP on $D$ if and only if $D$ is a \kalm{q} matrix.
\end{theorem}
\proof
For the if-part, we note that any principal sub-matrix of a \kalm{q} matrix $D$ again is a \kalm{q} matrix.
By Theorem~\ref{th:vlad} the identity permutation is an optimal solution for the $q$-stripe TSP on $D$ 
and induces optimal solutions for all principal sub-matrices.

For the only-if-part, we consider an arbitrary sequence of $2q+2$ cities $i_1<\cdots<i_{2q+2}$
in the considered instance.
These $2q+2$ cities span altogether $(q+1)(2q+1)$ edges.
Every $q$-stripe TSP tour uses exactly $q\,(2q+2)$ of these edges, and the remaining $q+1$ unused edges 
form a perfect matching.
As the identity permutation induces a minimum weight solution to the $q$-stripe TSP,
the unused edges (which form a fully crossing matching) should yield a matching of maximum weight.
This implies that the $2q+2$ cities satisfy the \kalm{q} condition.
\qed

\bigskip
By Theorem~\ref{th:master}, an instance of the $q$-stripe TSP possesses a master tour if and only
if the underlying distance matrix can be permuted into a \kalm{q} matrix.
At the current moment, we do not know whether it is easy or hard to recognize whether a given 
matrix can be permuted into a $q$-Kalmanson matrix. 
One might expect that the polynomial time algorithm of Deineko, Rudolf \& Woeginger \cite{DeRuWo1998}
for the special case $q=1$ could be extended to the cases with arbitrary $q$.
However, this is by no means not straightforward to do, as some of the combinatorial details in the
general case become quite complicated and messy.

\section{Hardness results}
\label{sec:hard}
In this section we return to the graph-theoretic version of the $q$-stripe TSP that we introduced
in Section~\ref{sec:literature}: We consider the problem of deciding the existence of a spanning 
sub-graph $C_n^q$ in a given undirected graph on $n$ vertices.
We show that this problem is hard in multi-partite graphs, in split graphs, and in graphs that do
not contain $K_{1,4}$ as induced sub-graph.
First, let us recall that a graph is $p$-partite if its vertex set can be partitioned into $p$ independent sets.
As the graph $C_n^q$ contains a complete sub-graph on $q+1$ vertices, the spanning sub-graph problem is trivial
(with a trivial negative answer) for all $p$-partite graphs with $p\le q$.
We will show that the problem is NP-hard even for $(q+1)$-partite graphs.
Next, let us recall that a split graph is a graph whose vertex set can be partitioned into one part 
that induces a clique and another part that induces an independent set.

The central hardness reduction is done from the following NP-complete HAMILTONIAN CIRCUIT problem; see 
Garey \& Johnson \cite{GaJo1979}.  

\begin{quote}
Problem: HAMILTONIAN CIRCUIT
\\[0.5ex]
Instance: A directed graph $G=(V,A)$.
\\[0.1ex]
Question: Does $G$ contain a (directed) Hamiltonian circuit?
\end{quote}
By definition every graph $C_n^q$ contains a Hamiltonian cycle with the following property: 
whenever two vertices are separated by at most $q-1$ vertices along the Hamiltonian cycle, then these 
vertices are also adjacent in $C_n^q$.
Such a Hamiltonian cycle will be called a \emph{Hamiltonian spine} of $C_n^q$.

\begin{figure}[bht]
\vspace{5ex}
\centerline{\includegraphics[width=14.2cm]{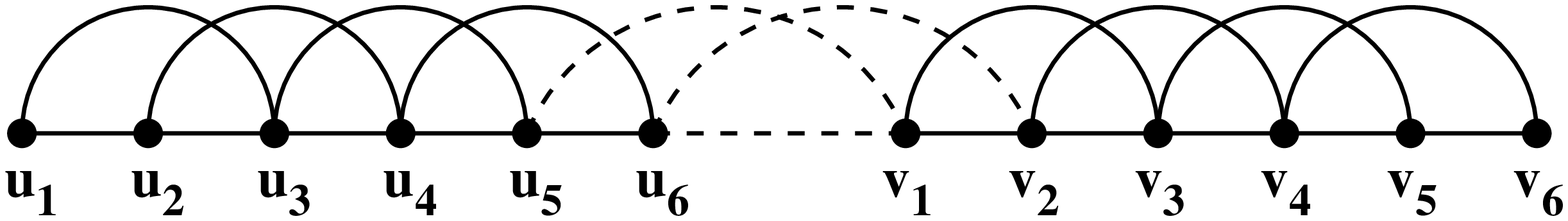}}
\caption{An illustration for the case $q=2$, showing the (solid) edges among $u_1,\ldots,u_6$ and 
 among $v_1,\ldots,v_6$ and the (dashed) edges connecting these two groups.}
\label{fig:circuit}
\end{figure}

We take an arbitrary instance $G=(V,A)$ of HAMILTONIAN CIRCUIT, and we construct the following undirected
graph $G_1=(V_1,E_1)$ from it.
For every vertex $v\in V$, the undirected graph $G_1$ contains $2q+2$ corresponding vertices
denoted $v_1,v_2,\ldots,v_{2q+2}$.
The edge set $E_1$ is defined as follows:
\begin{itemize}
\itemsep=0.0ex
\item For every $v\in V$, we create all the edges $[v_i,v_j]$ with $|i-j|\le q$.
\item For every arc $(u,v)\in A$, we create all the edges $[u_i,v_j]$ with $i-j\ge q+2$.
\end{itemize}
See Figure~\ref{fig:circuit} for an illustration.
For $\ell=0,\ldots,q$, define the vertex set $W_{\ell}$ to contain all vertices $v_i$ with $i\equiv\ell\bmod q+1$.
It is easily verified that graph $G_1$ is $(q+1)$-partite with partition $W_0,\ldots,W_q$.
Finally, we introduce the graph $G_2=(V_2,E_2)$ as a super-graph of $G_1$: the vertex set $V_2$ coincides with $V_1$, and
the edge set $E_2$ contains the edges in $E_1$ together with all edges on the vertex set $W_1\cup W_2\cup\cdots\cup W_q$.
Note that $G_2$ is a split graph with independent set $W_0$.

\begin{lemma}
\label{le:hard.1}
If the directed graph $G$ contains a Hamiltonian circuit, then the $(q+1)$-partite graph $G_1$ 
contains a spanning sub-graph $C_{nq}^q$.
\end{lemma}
\proof
Consider a Hamiltonian circuit in $G$, and replace every vertex $v$ in the circuit by the corresponding 
sequence $v_1,\ldots,v_{2q+2}$.
This yields the Hamiltonian spine for a spanning sub-graph $C_{nq}^q$ of $G_1$.
\qed

\begin{lemma}
\label{le:hard.2}
If the $(q+1)$-partite graph $G_1$ contains a spanning sub-graph $C_{nq}^q$,
then the split graph $G_2$ contains a spanning sub-graph $C_{nq}^q$.
\end{lemma}
\proof
The graph $G_2$ is a super-graph of $G_1$.
\qed

\begin{lemma}
\label{le:hard.3}
If the split graph $G_2$ contains a spanning sub-graph $C_{nq}^q$,
then the directed graph $G$ contains a Hamiltonian circuit. 
\end{lemma}
\proof
Consider the Hamiltonian spine of a spanning sub-graph $C_{nq}^q$ in $G_2$. 
The independent set $W_0$ contains $n$ vertices, and any two vertices in $W_0$ must be separated by at 
least $q$ other vertices along the Hamiltonian spine.
As the clique $W_1\cup W_2\cup\cdots\cup W_q$ contains only $nq$ vertices, this implies that along the
Hamiltonian spine any two consecutive vertices $x$ and $y$ from $W_0$ are separated by exactly $q$ vertices 
from the clique; each of these $q$ separating vertices is adjacent to both $x$ and $y$.

Now consider the $2q+2$ vertices $u_1,\ldots,u_{2q+2}$ that correspond to some fixed vertex $u\in V$.
As vertex $u_{q+1}$ has only $2q$ neighbors in the split graph (the vertices $u_1,\ldots,u_q$ and 
$u_{q+2},\ldots,u_{2q+1}$), some $q$ of these neighbors must directly precede $u_{q+1}$ in the Hamiltonian 
spine while the other $q$ neighbors must succeed it.
We assume without loss of generality that the neighbor $u_{q+2}$ is among the $q$ vertices that succeed $u_{q+1}$.
As vertex $u_{q+2}$ has only two neighbors in $W_0$ (the vertices $u_{q+1}$ and $u_{2q+2}$), this means that 
along the Hamiltonian spine vertex $u_{2q+2}$ is the first $W_0$-vertex after $u_{q+1}$.
All in all, this means that the $2q+2$ vertices $u_1,\ldots,u_{2q+2}$ occur as a single block along the spine
with $u_{2q+2}$ occurring as last vertex.

Next consider the vertex $v_i$ that directly follows $u_{2q+2}$ in the Hamiltonian spine.
Then $[u_{2q+2},v_i]\in E_2$, which implies that $(u,v)$ is an arc in the directed graph $G$.
Similarly as in the preceding paragraph it can furthermore be seen that also the $2q+2$ vertices 
$v_1,\ldots,v_{2q+2}$ occur as a single block along the spine with $v_{2q+2}$ occurring as last vertex.
Finally, a simple inductive argument shows that replacing every group $v_1,\ldots,v_{2q+2}$ in the spine 
by their corresponding vertex $v\in V$ yields a Hamiltonian circuit for the directed graph $G$.
\qed

\bigskip
By combining Lemmas~\ref{le:hard.1}, \ref{le:hard.2} and \ref{le:hard.3}, 
we derive the following theorem.

\begin{theorem}
\label{th:np.1}
For every $q\ge2$, it is NP-complete to decide whether (i) a given $(q+1)$-partite graph and (ii)
a given split graph contains a spanning $q$-stripe tour.
\qed
\end{theorem}

\begin{theorem}
\label{th:np.2}
For every $q\ge2$, it is NP-complete to decide whether a graph without induced sub-graph $K_{1,4}$ contains 
a spanning $q$-stripe tour.
\end{theorem}
\proof
Plesnik \cite{Plesnik1979} has shown that the HAMILTONIAN CIRCUIT problem is NP-complete, even if the 
underlying directed graph $G=(V,A)$ has in-degrees and out-degrees at most~$2$.
We claim that if we start the above reduction from such a directed graph $G$, then the resulting split 
graph $G_2$ does not contain $K_{1,4}$ as induced sub-graph.

Suppose otherwise, and consider the central vertex $x$ of an induced $K_{1,4}$ in $G_2$.
Then $x$ must be in the clique, and its four neighbors must be in the independent set $W_0$.
Hence each of the four neighbors must be a vertex $u_{q+1}$ or $u_{2q+2}$ for some $u\in V$.
\begin{itemize}
\item If $u_{q+1}$ is one of these four neighbors, then $x$ is among the $2q$ vertices 
$u_1,\ldots,u_q$ and $u_{q+2},\ldots,u_{2q+1}$.
\item If $u_{2q+2}$ is one of these four neighbors, then $x$ is among the vertices 
$u_{q+2},\ldots,u_{2q+1}$ or among the vertices $v_1,\ldots,v_q$ for some $v\in V$ with $(u,v)\in A$.
\end{itemize}
If $x$ is among vertices $u_{q+2},\ldots,u_{2q+1}$ for some $u\in V$, then $x$ has only two
possible neighbors in $W_0$ (the two vertices $u_{q+1}$ and $u_{2q+2}$).
Hence $x$ must be one of $u_1,\ldots,u_q$ for some $u\in V$.
Then $x$ has at most three neighbors in $W_0$: the vertex $u_{q+1}$, and perhaps two vertices
$v_{2q+2}$ and $w_{2q+2}$ with $(v,u)\in A$ and $(w,u)\in A$.
\qed

\bigskip
Theorems~\ref{th:np.1} and \ref{th:np.2} also imply the following negative result on the $q$-stripe TSP.  
\begin{corollary}
\label{co:np.1a}
For every $q\ge2$, the $q$-stripe TSP is NP-complete even if the distance matrix is a symmetric $0$-$1$ matrix.
\qed
\end{corollary}

Another immediate consequence of Theorems~\ref{th:np.1} and \ref{th:np.2} concerns the bottleneck version 
of the multi-stripe TSP, where the objective is to minimize the length of the longest used edge (instead 
of minimizing the total sum of all used edges).

\begin{corollary}
\label{co:np.1b}
For every $q\ge2$, the bottleneck version of the $q$-stripe TSP is NP-complete.
\end{corollary}

\section{Polynomial time results}
\label{sec:easy}
In this section we discuss the problem of finding spanning $q$-stripe tours in planar graphs and in partial $k$-trees.
Let us start with the problem of deciding the existence of a spanning $q$-stripe tour $C_n^q$ in a given
planar graph $G=(V,E)$ on $n$ vertices.
For $q=1$, this decision problem is the standard Hamiltonian cycle problem and hence NP-complete for 
planar graphs; see Garey \& Johnson \cite{GaJo1979}.
On the other hand for $q\ge3$ this problem is trivial: Every planar graph contains a vertex of degree 
at most $5$, whereas the graph $C_n^q$ is $2q$-regular; hence for $q\ge3$ the answer will always be negative.
Summarizing, the only interesting version of this spanning sub-graph problem is the case with $q=2$.

\begin{lemma}
\label{le:planar.1}
For $n\ge5$ the graph $C_n^2$ is planar, if and only if $n$ is even.
\end{lemma}
\proof
For even $n$, the graph $C_n^2$ decomposes into three cycles: the Hamiltonian spine $1,2,\ldots,n$;
a cycle of length $n/2$ traversing the even vertices $2,4,6,\ldots,n$;
and a cycle of length $n/2$ traversing the odd vertices $1,3,5,\ldots,n-1$.
The Hamiltonian spine is easily embedded and partitions the plane into a bounded face and an unbounded face.
We embed the edges of the cycle $2,4,6,\ldots,n$ in the bounded face, and
we embed the edges of the cycle  $1,3,5,\ldots,n-1$ in the unbounded face.
Hence $C_n^2$ is planar.

For odd $n$, we observe that $C_5^2$ is the non-planar complete graph on five vertices.
For $n\ge7$, graph $C_n^2$ contains a subdivision of the (non-planar) complete bipartite graph $K_{3,3}$.
Indeed, we may embed one side of the bipartition into the vertices $1,4,5$ and the other side into $2,3,6$.
Then the seven edges $[1,2]$, $[1,3]$, $[4,2]$, $[4,3]$, $[4,6]$, $[5,3]$, and $[5,6]$ are contained in $C_n^2$.
The edge $[1,6]$ results from the path $6-8-10-\cdots-(n-1)-1$.
Finally the edge $[5,2]$ results from the path $5-7-9-\cdots-n-2$.
\qed

\begin{lemma}
\label{le:planar.2}
Let $G$ be a planar graph that contains five vertices $u,v,x,y,z$ so that $u$ and $v$ are adjacent, 
and so that $x,y,z$ are common neighbors of both $u$ and $v$.
Then $G$ does not contain $C_n^2$ as a spanning sub-graph.
\end{lemma}
\proof
If $G$ contains such five vertices $u,v,x,y,z$, in any planar embedding one of the three triangles $u,v,x$ 
and $u,v,y$ and $u,v,z$ will be a separating triangle for $G$.
Thus $G$ has a $3$-element cut set, whereas the graph $C_n^2$ does not allow such a cut set.
\qed

\bigskip
Now suppose that some planar graph $G$ contains a spanning sub-graph $C_n^2$.
Let $v_1,v_2,\ldots,v_n$ be the underlying Hamiltonian spine, so that any two vertices $v_i$ and $v_j$ 
with $|i-j|\le2$ or $|i-j|\ge n-2$ are adjacent in $G$.
We claim that the first three vertices $v_1,v_2,v_3$ in the spine already determine the full spanning 
sub-graph $C_n^2$.
Indeed, the three vertices $v_1,v_2,v_3$ then induce a triangle, and any candidate for the fourth vertex 
in the spine must be adjacent to both $v_2$ and $v_3$.
If there were two distinct candidates for the fourth vertex, then these two candidates together with 
$v_1,v_2,v_3$ would yield the forbidden configuration in Lemma~\ref{le:planar.2}.
Hence there is at most one candidate for the fourth vertex.
Arguing inductively, this fully determines the spine and hence the spanning sub-graph $C_n^2$.
By trying all possibilities for $v_1,v_2,v_3$, this leads to the following theorem.

\begin{theorem}
\label{th:planar}
It can be decided in polynomial time whether a given planar graph on $n$ vertices contains a 
spanning sub-graph $C_n^2$.
\qed
\end{theorem}

Finally let us turn to partial $k$-trees, which form a well-known generalization of ordinary trees; 
see for instance the survey articles by Bodlaender \cite{Bod1,Bod2,Bod3} for more information.
We mention as an example that series-parallel graphs and outerplanar graphs are partial $2$-trees.
Many algorithmic problems can be solved in polynomial time on partial $k$-trees, as long as the
value $k$ is constant and not part of the input.
More precisely, every graph problem that is expressible in Monadic Second Order Logic (MSOL) is solvable 
in linear time on partial $k$-trees with constant $k$; see Arnborg, Lagergren \& Seese \cite{ArLaSe1991}.

\begin{theorem}
\label{th:partial}
For every $q\ge2$ and for every $k\ge1$, it can be decided in linear time whether a given partial 
$k$-tree contains a spanning sub-graph $C_n^q$.
\end{theorem}
\proof
For a given graph $G=(V,E)$, the property of having a spanning sub-graph $C_n^q$ 
can be expressed in MSOL as follows:
\begin{itemize} 
\item There exists a set $F\subseteq E$, so that every vertex is incident to exactly two edges in $F$.
\item There does not exist any partition of the vertex set $V$ into two non-empty sets $V_1$ and $V_2$, 
so that none of the edges in $F$ connects $V_1$ to $V_2$.
\item For any sequence $v_1,v_2,\ldots,v_r$ of $r\le q$ vertices: if $[v_s,v_{s+1}]\in F$ for $1\le s\le r-1$,
then $[v_1,v_r]\in E$.  
\end{itemize}
Each of these statements can be formulated in MSOL in a straightforward way.
The first two statements make $F$ the edge set of a Hamiltonian spine.
The third statement ensures that all edges in $C_n^q$ outside the spine are also present in graph $G$.
\qed

\section{Discussion}
\label{sec:discussion}
We have derived a number of positive and negative results on the $q$-stripe TSP.
As our main result, we have introduced the class of $q$-Kalmanson matrices and we have generalized a 
well-known result of Kalmanson on the classical TSP to the $q$-stripe TSP on matrices from this class.
As a by-product, our investigations yield a complete analysis of the so-called master version of the
$q$-stripe TSP, where the master solution simultaneously induces optimal solutions to all possible
sub-instances of a given problem instance.
Furthermore, we have analyzed the graph-theoretic version of the $q$-stripe TSP.
We derived NP-completeness for $(q+1)$-partite graphs and for split graphs, and we derived polynomial
time results for planar graphs (if $q\ge2$) and for partial $k$-trees (if $k$ is a fixed constant).

There are many open questions around the $q$-stripe TSP.
First of all, we would like to understand the $q$-stripe TSP on so-called Demidenko matrices.
An $n\times n$ matrix $D$ is a \emph{Demidenko matrix} if its entries fulfill the conditions \eqref{ka.1}.
A celebrated result \cite{Demidenki1979} of Demidenko (see also Gilmore, Lawler \& Shmoys \cite{GLS1985})
shows that the classical TSP on Demidenko matrices is solvable in polynomial time. 
We did not manage to settle the complexity of the $q$-stripe TSP on Demidenko matrices, and even the 
case $q=2$ is unclear.
It might well be possible that this problem turns out to be NP-hard.

Deineko, Klinz, Tiskin \& Woeginger \cite{DeKlTiWo2014} analyze the classical TSP with respect
to so-called \emph{four-point conditions} on the distance matrix, that is, constraining inequalities 
that involve the distances between four arbitrary cities.
For instance, Monge matrices, Kalmanson matrices, and Demidenko matrices fall under this framework. 
Furthermore, there are $18$ other natural classes of distance matrices in the framework, and some of
these classes might allow interesting results for the $q$-stripe TSP.

Also the graph-theoretic version of the $q$-stripe TSP is quite poorly understood, and the computational
complexity is open for many natural graph classes.
Our hardness result for split graphs trivially yields hardness for the broader class of chordal graphs,
and of course for the class of perfect graphs.
But for other classes of perfect graphs, as for instance for permutation graphs and for strongly chordal 
graphs, the complexity of the $q$-stripe TSP remains unclear.
In particular the complexity is open for interval graphs; see Isaak \cite{Isaak1998}.
We have shown in Section~\ref{sec:hard} that the $q$-stripe TSP is NP-hard on graphs without induced 
sub-graph $K_{1,4}$.
We note that the complexity for claw-free graphs (that is, for graphs that do not contain $K_{1,3}$ 
as induced sub-graph) is open for $q\ge2$.
The classical case with $q=1$ is known to be NP-complete; see Bertossi \cite{Bertossi1981}.

\medskip
{\small
\paragraph{Acknowledgements.}
This research was conducted while Vladimir Deineko and Gerhard Woeginger were visiting TU Graz,
and they both thank the Austrian Science Fund (FWF): W1230, Doctoral Program in ``Discrete Mathematics'' 
for the financial support.
Vladimir Deineko acknowledges support
  by Warwick University's Centre for Discrete Mathematics and Its Applications (DIMAP).
Gerhard Woeginger acknowledges support
  by the Zwaartekracht NETWORKS grant of NWO.}


\end{document}